\documentclass[12pt]{article}
\usepackage{amssymb,amsmath,epsfig}

\numberwithin{equation}{section}

\begin{document}

\title{\textbf{Fermions Tunneling from Charged Accelerating
and Rotating Black Holes with NUT Parameter}}

\author{M. Sharif \thanks{msharif.math@pu.edu.pk} and Wajiha Javed
\thanks{wajihajaved84@yahoo.com} \\
Department of Mathematics, University of the Punjab,\\
Quaid-e-Azam Campus, Lahore-54590, Pakistan.}

\date{}
\maketitle

\begin{abstract}
This paper is devoted to the study of Hawking radiation as a
tunneling of charged fermions through event horizons of a pair of
charged accelerating and rotating black holes with NUT parameter. We
evaluate tunneling probabilities of outgoing charged particles by
using the semiclassical WKB approximation to the general covariant
Dirac equation. The Hawking temperature corresponding to this pair
of black holes is also investigated. For the zero NUT parameter, we
find results consistent with those already available in the
literature.
\end{abstract}

\section{Introduction}

\emph{Hawking radiation} \cite{hawk1} is a quantum mechanical
process which can be described as follows: Firstly, a pair of
positive and negative energy particles is created near the horizon.
The negative energy particle falls inside the horizon, while the
positive energy particle tunnels outside the horizon and rises in as
Hawking radiation for the outside observer. Secondly, the pair of
particles is created just inside the horizon. The positive energy
particles have ability to cross the energy barrier and tunnels to
infinity, however the negative energy partner remains behind. This
reduces the mass of black hole (BH) by losing energy in the form of
Hawking radiation. The process in which particles have finite
probability to cross the event horizon (that cannot be possible for
classical particles) is called \emph{quantum tunneling}. In this
process, we evaluate tunneling probabilities that depend on the
imaginary part of the action for emitted particles through the
horizon.

There are two methods to find the imaginary part of the action for
the emitted particles. The first method was developed by Parikh and
Wilczek \cite{parwil} named as the \emph{null geodesic method} by
following the Kraus and Wilczek \cite{krawil} technique. The second
one is called \emph{Hamilton$-$Jacobi method} which was first used
by Angheben \emph{et al} \cite{abhen} for tunneling phenomenon to
extend the complex path analysis of Srinivasan \emph{et al}
\cite{Srini1,Srini2}.  The tunneling probability $\Gamma$ for the
emitted particle can be written as
\begin{equation}
\Gamma\propto\exp[-2\textmd{Im}I],
\end{equation}
where $I$ is the classical action ($\hbar=1$). An approximation,
called \emph{WKB approximation} was used by Wentzel, Kramers and
Brillouin (WKB) \cite{WKB} to develop a relation between classical
and quantum theory. This is used to solve the Schr$\ddot{o}$dinger
equation ($\hbar=0$ leads to Hamilton$-$Jacobi equation in classical
theory) in quantum mechanics. In both the above methods, the WKB
approximation is used.

The study of Hawking radiation as a tunneling phenomenon from
various BHs has attracted many people. Page \cite{23-1} calculated
the emission rates from an uncharged, nonrotating BH for massless
particles of spin $\frac{1}{2},~1$ and $2$ by using the perturbation
formalism. The same author \cite{23-2} found numerical calculations
of the emission rates of massless particles from a rotating BH.
Kerner and Mann \cite{rbmann1} extended the calculations of the
tunneling process for the spin $\frac{1}{2}$ particles emission by
using the WKB approximation to the Dirac equation and found the
tunneling probability from nonrotating BHs. Also, fermions tunneling
is applied to a general nonrotating BH and recovered the
corresponding Hawking temperature. The same authors \cite{rbmann}
also discussed the tunneling of charged spin $\frac{1}{2}$ fermions
from a Kerr$-$Newman BH and obtained the Hawking temperature.

In recent papers, the tunneling probabilities of incoming and
outgoing scalar and charged/uncharged fermions from accelerating and
rotating BHs have been investigated \cite{ks1}. Also, the
thermodynamical properties of accelerating and rotating BHs with
Newman$-$Unti$-$Tamburino (NUT) parameter have been studied
\cite{1-21}. Recently, we have also explored some work \cite{S1}
about the tunneling phenomenon for different BHs by using the above
mentioned methods. In this paper, we extend this tunneling
phenomenon of charged fermions from a pair of charged accelerating
and rotating BHs with NUT parameter by using the procedure
\cite{rbmann1}.

The paper is outlined as follows. In the next section, we briefly
review Hamilton$-$Jacobi ansatz. Section \textbf{3} is devoted to
explain the basic equations for a pair of accelerating and rotating
BHs with NUT parameter having electric and magnetic charges. In
section \textbf{4}, we deal with the solution of the Dirac equation
of charged particles in the background of these BHs and evaluate the
tunneling probabilities as well as the corresponding temperature
across the horizons. Section \textbf{5} provides an exact form of
the action for the massless and massive particles. Finally, we
summarize the results in the last section.

\section{Hamilton$-$Jacobi Ansatz: An Overview}

In this section, we briefly review Hawking radiation as tunneling
through Hamilton$-$Jacobi ansatz \cite{krawil, abhen, rbmann1}. The
metric for a general (non-extremal) BH is
\begin{equation}
ds^2=-f(r)dt^2+\frac{dr^2}{g(r)}+C(r)h_{ij}dx^idx^j.\label{W-W}
\end{equation}
Assume a scalar particle moving in this BH. Since we are interested
only in the leading terms within the semiclassical approximation, we
neglect the effects of the particle self-gravitation. Thus the
quantity which satisfies the relativistic Hamilton$-$Jacobi equation
is the classical action $I$
\begin{equation}
g^{\mu\nu}\partial_\mu I\partial_\nu I+m^2=0
\end{equation}
which implies that
\begin{equation}
-\frac{(\partial_tI)^2}{f(r)}+g(r)(\partial_r
I)^2+\frac{h^{ij}}{C(r)}\partial_iI\partial_jI+m^2=0.
\end{equation}
Due to the symmetries of the metric, there exists a solution of the
form
\begin{equation}\label{a}
I=-Et+W(r)+J(x^i),
\end{equation}
where $\partial_t I =-E,~ \partial_r I=W^\prime(r),~ \partial_i
I=J_i$ and the $J_i$'s are constants. $E$ corresponds to energy as
$\partial_t$ is the timelike Killing vector. Solving for $W(r)$, it
follows that
\begin{equation}
W_\pm(r)=\pm\int\frac{dr}{\sqrt{f(r)g(r)}}\sqrt{E^2-f(r)
\left(m^2+\frac{h^{ij}J_iJ_j}{C(r)}\right)}.
\end{equation}
Here, $W_+$ and $W_-$ correspond to scalar particles moving away
(outgoing) and moving towards the BH (incoming), orespectively.
Notice that the imaginary part of the action can only be due to the
pole at the horizon.

The probabilities of crossing the horizon each way are
\begin{eqnarray}
\textmd{Prob}[\textmd{out}]\propto\exp[-\frac{2}{\hbar}\textmd{Im}I]=
\exp[-\frac{2}{\hbar}\textmd{Im}W_+],\\
\textmd{Prob}[\textmd{in}]\propto\exp[-\frac{2}{\hbar}\textmd{Im}I]=
\exp[-\frac{2}{\hbar}\textmd{Im}W_-].
\end{eqnarray}
Since $W_+=-W_-$, we obtain the probability of a particle tunneling
from inside to outside the horizon
\begin{equation}
\Gamma\propto\exp[-\frac{4}{\hbar}\textmd{Im}W_+].
\end{equation}
We take $\hbar$ to be unity and also drop the '+' subscript from
$W$. Integrating around the pole at the horizon leads to the result
\cite{s2}
\begin{equation}
W=\frac{\pi\iota E}{\sqrt{f^\prime(r_0)g^\prime(r_0)}}.
\end{equation}
Consequently, the tunneling probability becomes
\begin{equation}
\Gamma=\exp[-\frac{4\pi}{\sqrt{f^\prime(r_0)g^\prime(r_0)}}E]
\end{equation}
which leads to the usual Hawking temperature
\begin{equation}
T_H=\frac{\sqrt{f^\prime(r_0)g^\prime(r_0)}}{4\pi}.
\end{equation}

\section{Accelerating and Rotating NUT Solutions}

Black holes are the most important predictions of general
relativity. The discovery of Schwarzschild BH is followed by the
extension to the electrically charged version,
Reissner$-$Nordstr\"{o}m (RN) BH. This is followed by the discovery
of rotating versions, Kerr and Kerr$-$Newman BHs \cite{[1]}. The
research in the BH area has further been extended by adding
different sources like cosmological constant, an acceleration as
well as a NUT parameter. Black hole solutions with these extensions
belong to type-D class \cite{[8]}.

In general, the NUT parameter is associated with the gravitomagnetic
monopole parameter of the central mass, or a twisting property of
the surrounding spacetime but its exact physical meaning could not
be ascertained. Recently, higher dimensional generalization of the
Kerr$-$NUT$-$(anti) de Sitter spacetime \cite{[9]} and its physical
significance \cite{[11]} is investigated. For this BH, the dominance
of the NUT parameter over the rotation parameter leaves the
spacetime free of curvature singularities and the corresponding
solution is named as NUT-like solution. However, if the rotation
parameter dominates the NUT parameter, the solution is Kerr-like and
a ring curvature singularity forms. This kind of behavior on the
singularity structure is independent of the presence of the
cosmological constant.

There are many BH solutions which incorporate the NUT parameter and
investigate its physical effect in the space of colliding waves.
Exact interpretation of the NUT parameter becomes possible when a
static Schwarzschild mass is immersed in a stationary, source free
electromagnetic universe \cite{[12]}. In this case, the NUT
parameter is related to the twist of the electromagnetic universe
excluding the central Schwarzschild mass. In the absence of this
field, it reduces to the twist of the vacuum spacetime. Thus, the
twist of the environmental space couples with the mass of the source
to generate NUT parameter. It would be interesting to explore the
tunneling phenomenon from such a BH.

The family of solutions which represents a pair of charged
accelerating and rotating BHs with a nonzero NUT parameter can be
written as \cite{acc}
\begin{eqnarray}
\textmd{d}s^2&=&-\frac{1}{\Omega^2}\left\{\frac{Q}{\rho^2}
\left[\textmd{d}t-\left(a\sin^2\theta+4l\sin^2\frac{\theta}{2}
\right)\textmd{d}\phi\right]^2- \frac{\rho^2}{Q}
\textmd{d}r^2\right.\nonumber\\
&-&\left.\frac{\tilde{P}}{\rho^2}\left[a\textmd{d}t-\left(r^2+(a+l)^2\right)
\textmd{d}\phi\right]^2-\frac{\rho^2}{\tilde{P}}\sin^2\theta
\textmd{d}\theta^2\right\}\label{NN},
\end{eqnarray}
where
\begin{eqnarray}
\Omega&=&1-\frac{\alpha}{\omega}(l+a\cos\theta)r,\quad
\rho^2=r^2+(l+a\cos\theta)^2,\nonumber\\
Q&=&\left[(\omega^2k+e^2+g^2)(1+2\frac{\alpha l
}{\omega}r)-2Mr+\frac{\omega^2k}{a^2-l^2}r^2\right]\nonumber\\
&\times&\left[1+\frac{\alpha(a-l)}{\omega}r\right]
\left[1-\frac{\alpha(a+l)}{\omega}r\right],\nonumber\\
\tilde{P}&=&\sin^2\theta(1-a_3\cos\theta-a_4\cos^2\theta)=P\sin^2\theta,\nonumber\\
a_3&=&2\frac{\alpha
a}{\omega}M-4\frac{\alpha^2 al}{\omega^2}(\omega^2k+e^2+g^2),\quad
a_4=-\frac{\alpha^2 a^2}{\omega^2}(\omega^2k+e^2+g^2).\nonumber
\end{eqnarray}
Here $M$ represents a pair of BHs mass, $e$ and $g$ indicate
electric and magnetic charges, respectively, while $a$ shows a BH
rotation and $l$ is a NUT parameter \cite{NUT}. The continuous
parameter $\alpha$ represents acceleration of the sources. Also, the
rotation parameter $\omega$ is related to the rotation of the
sources and $k$ is given by
\begin{equation}
\left(\frac{\omega^2}{a^2-l^2}+3\alpha^2l^2\right)k=1+2\frac{\alpha
l}{\omega}M-3\frac{\alpha^2l^2}{\omega^2}(e^2+g^2),\label{WWW}
\end{equation}
where $\alpha,~\omega,~M,~e,~g$ and $k$ are arbitrary real
parameters.

Notice that this type of BH involves acceleration $\alpha$ while
twisting behavior of the sources is proportional to the parameter
$\omega$ which is related to both the Kerr-like rotation parameter
$a$ and the NUT parameter $l$. Thus, $\alpha,~M,~e,~g,~l,a$ vary
independently while $\omega$ depends on nonzero value of rotation
parameters $l$ or $a$. For $\alpha=0$, this reduces to the
Kerr$-$Newman$-$NUT solution and Eq.(\ref{WWW}) will become
$\omega^2k=a^2-l^2$. In the absence of NUT parameter $l=0$, it
reduces to the pair of charged accelerating and rotating BHs.
Further, $\alpha=0$ leads to the Kerr$-$Newman BH and $a=0$ yields
the RN BH. In addition, if $e=0=g$, we have a Schwarzschild BH while
$l=0=a$ leads to the C-metric.

The metric (\ref{NN}) can also be written in a more suitable form
as
\begin{equation}
\textmd{d}s^2=-f(r,\theta)\textmd{d}t^2
+\frac{\textmd{d}r^2}{g(r,\theta)}+\Sigma(r,\theta)
\textmd{d}\theta^2+K(r,\theta)\textmd{d}
\phi^2-2H(r,\theta)\textmd{d}t\textmd{d}\phi,
\end{equation}
where $f(r,\theta),~g(r,\theta),~\Sigma(r,\theta),~K(r,\theta)$ and
$H(r,\theta)$ can be defined as follows:
\begin{eqnarray}
f(r,\theta)&=&\left(\frac{Q-Pa^2\sin^2\theta}{\rho^2\Omega^2}
\right),\quad g(r,\theta)=\frac{Q\Omega^2}
{\rho^2},\quad\Sigma(r,\theta)=\frac{\rho^2}{\Omega^2P},\\
K(r,\theta)&=&\frac{1}{\rho^2\Omega^2}\left[\sin^2\theta
P[r^2+(a+l)^2]^2-Q(a\sin^2\theta+4l\sin^2
\frac{\theta}{2})^2\right],\\
H(r,\theta)&=&\frac{1}{\rho^2\Omega^2}\left[\sin^2\theta
Pa[r^2+(a+l)^2]-Q(a\sin^2 \theta+4l\sin^2\frac{\theta}{2})\right].
\end{eqnarray}
The electromagnetic vector potential for these BHs is given by
\cite{acc1}
\begin{eqnarray}
A&=&\frac{1}{a[r^2+(l+a\cos\theta)^2]}
[-er[a\textmd{d}t-\textmd{d}\phi\{(l+a)^2
-(l^2+a^2\cos^2\theta\nonumber\\&+&2la\cos\theta)\}]
-g(l+a\cos\theta)[a\textmd{d}t-\textmd{d}\phi\{r^2+(l+a)^2\}]].
\end{eqnarray}

The event horizons are obtained for
$g(r,\theta)=\frac{\Delta(r)}{\Sigma(r,\theta)}=0$ \cite{rbmann},
where $\Delta(r)=\frac{Q}{P}$. This implies that $\Delta(r)=0=Q$,
yielding the horizon radii
\begin{eqnarray}
r_{\alpha_1}=\frac{\omega}{\alpha(a+l)},\quad
r_{\alpha_2}=-\frac{\omega} {\alpha(a-l)},\quad
r_\pm=\frac{a^2-l^2}{\omega^2k}
\left[-[(\omega^2k+e^2+g^2)\frac{\alpha
l}{\omega}\right.\nonumber\\\left.-M]\pm\sqrt{[(\omega^2
k+e^2+g^2)\frac{\alpha l}{\omega}-M]^2-\frac{\omega^2k}
{a^2-l^2}(\omega^2k+e^2+g^2)}\right],\label{11}
\end{eqnarray}
where $r_\pm$ represent the outer and inner horizons, respectively,
such that $[(\omega^2 k+e^2+g^2)\frac{\alpha
l}{\omega}-M]^2-\frac{\omega^2k}{a^2-l^2}(\omega^2k+e^2+g^2)>0$,
$r_{\alpha_1}$ and $r_{\alpha_2}$ are acceleration horizons. The
angular velocity at the BH horizon can be defined as
\begin{equation}
\Omega_H=\frac{H(r_+,\theta)}{K(r_+,\theta)}=\frac{a}{r_+^2+(a+l)^2}.
\end{equation}
The inverse function of $f(r,\theta)$ is
\begin{equation}
F(r,\theta)=f(r,\theta)+\frac{H^2(r,\theta)}{K(r,\theta)}.
\end{equation}
For these BHs, this can be written in the following form:
\begin{equation}
F(r,\theta)=\frac{PQ\sin^2\theta\rho^2}{\Omega^2[\sin^2\theta
P[r^2+(a+l)^2]^2-Q(a\sin^2\theta+4l\sin^2\frac{\theta}{2})^2]}.
\end{equation}
In terms of $\Delta(r)$ and $\Sigma$, we can write the inverse
function of $f(r,\theta)$ as
\begin{equation}
F(r,\theta)=\frac{P^2\Delta(r)\Sigma(r,\theta)}{[r^2+(a+l)^2]^2
-\Delta(r)\sin^2\theta[a+\frac{2l}{1+\cos\theta}]^2}.
\end{equation}

\section{Tunneling of Charged Fermions}

In order to study charged fermions tunneling of mass $m$ from a pair
of accelerating and rotating BHs with NUT parameter having electric
and magnetic charges, the covariant Dirac equation with electric
charge $q$ is given by \cite{rn}
\begin{equation}
\iota\gamma^\mu\left(D_\mu-\frac{\iota
q}{\hbar}A_\mu\right)\Psi+\frac{m}{\hbar}\Psi=0,\quad
\mu=0,1,2,3\label{2'}
\end{equation}
where $A_\mu$ is the 4-potential, $\Psi$ is the wave function and
\begin{equation}
D_\mu=\partial_\mu+\Omega_\mu,\quad
\Omega_\mu=\frac{1}{2}\iota\Gamma^{\alpha\beta}_
\mu\Sigma_{\alpha\beta},\quad\Sigma_{\alpha\beta}=
\frac{1}{4}\iota[\gamma^\alpha,\gamma^\beta].
\end{equation}
The antisymmetric property of the Dirac matrices \cite{ks1}, i.e.,
$[\gamma^\alpha,\gamma^\beta]=0$ for $\alpha=\beta$ and
$[\gamma^\alpha,\gamma^\beta]=-[\gamma^\beta,\gamma^\alpha]$ for
$\alpha\neq\beta$, reduces the Dirac equation (\ref{2'}) in the
following form:
\begin{equation}
\iota\gamma^\mu\left(\partial_\mu-\frac{\iota
q}{\hbar}A_\mu\right)\Psi+\frac{m}{\hbar}\Psi=0.\label{2}
\end{equation}
The Dirac matrices are given by
\begin{eqnarray}
\gamma^t&=&\frac{1}{\sqrt{F(r,\theta)}}\left(\begin{array}{cc}
\iota &0\\
0 &-\iota\\
\end{array}\right),\quad \gamma^r=\sqrt{g(r,\theta)}
\left(\begin{array}{cc}
0 &\sigma^3\\
\sigma^3 &0\\
\end{array}\right),\nonumber\\ \gamma^\theta&=&\frac{1}
{\sqrt{\Sigma(r,\theta)}}\left(\begin{array}{cc}
0 &\sigma^1\\
\sigma^1 &0\\
\end{array}\right),\nonumber\\
\gamma^\phi&=&\frac{1}{\sqrt{K(r,\theta)}}\left
[\left(\begin{array}{cc}
0 &\sigma^2\\
\sigma^2 &0\\
\end{array}\right)\nonumber\right.+\left.\frac{H(r,\theta)}{\sqrt{F(r,\theta)
K(r,\theta)}}\left(\begin{array}{cc}
\iota &0\\
0 &-\iota\\
\end{array}\right)\right],\label{}
\end{eqnarray}
where $\sigma^i~(i=1,2,3)$ are the Pauli sigma matrices defined as
\begin{equation}
\sigma^1=\left(\begin{array}{cc}
0 &1\\
1 &0\\
\end{array}\right),\quad \sigma^2=\left(\begin{array}{cc}
0 &-\iota\\
\iota &0\\
\end{array}\right),\quad \sigma^3=\left(\begin{array}{cc}
1 &0\\
0 &-1\\
\end{array}\right).\label{}
\end{equation}
The spinor wave function $\Psi$ (related to the particle's action)
has two spin states: spin-up (in +ve $r$-direction) and spin-down
(in -ve $r$-direction). For the spin-up and spin-down particle's
solution, we assume \cite{rbmann1}
\begin{eqnarray}
\Psi_\uparrow(t,r,\theta,\phi)&=&\left[\begin{array}{c}
A(t,r,\theta,\phi)\\0\\
B(t,r,\theta,\phi)\\0
\end{array}\right]\exp\left[\frac{\iota}{\hbar}
I_\uparrow(t,r,\theta,\phi)
\right],\label{1}\\
\Psi_\downarrow(t,r,\theta,\phi)&=&\left[\begin{array}{c}
0\\C(t,r,\theta,\phi)\\0\\
D(t,r,\theta,\phi)
\end{array}\right]\exp\left[\frac{\iota}{\hbar}
I_\downarrow(t,r,\theta,\phi) \right],\label{}
\end{eqnarray}
where $I_{\uparrow/\downarrow}$ denote the emitted spin-up/spin-down
particle's action, respectively. Note that we shall only analyze the
spin-up case as the spin-down case is just analogous.

The particle's action is given by the following ansatz (\ref{a}):
\begin{equation}
I_\uparrow=-Et+J\phi+W(r,\theta),\label{WWWW}
\end{equation}
where $E,~J$ and $W$ are energy, angular momentum and arbitrary
function, respectively. Using this ansatz into the Dirac equation
with $\iota A=B,~\iota B=A$ and Taylor's expansion of $F(r,\theta)$
near the event horizon, it follows that
\begin{eqnarray}
&-&B\left[\frac{-E+\Omega_HJ+\frac{qer}{[r_+^2+(a+l)^2]}}
{\sqrt{(r-r_+)\partial_rF(r_+,\theta)}}
+\sqrt{(r-r_+)\partial_rg(r_+,\theta)}(\partial_rW)\right]+mA=0,\nonumber\\\label{3}\\
&-&B\left[\sqrt{\frac{P\Omega^2
(r_+,\theta)}{\rho^2(r_+,\theta)}} (\partial_\theta W)\nonumber\right.\\
&+&\left.\frac{\iota\rho(r_+,\theta)
\Omega(r_+,\theta)}{\sqrt{\sin^2\theta
P[r^2+(a+l)^2]^2-Q(a\sin^2\theta+4l\sin^2\frac{\theta}{2})^2}}\right.\nonumber\\
&\times&\left.\left\{J-q
\left[\frac{er[(l+a)^2-(l^2+a^2\cos^2\theta+2la\cos\theta)]}
{a[r^2+(l+a\cos\theta)^2]}\right.
\right.\right.\nonumber\\
&+&\left.\left.\left.\frac{g(l+a\cos\theta)[r^2+(l+a)^2]}{a[r^2+(l+a\cos\theta)^2]}
\right]\right\}\right]=0,\label{4}
\end{eqnarray}
\begin{eqnarray}
&&A\left[\frac{-E+\Omega_HJ+\frac{qer}{[r^2+(a+l)^2]}}
{\sqrt{(r-r_+)\partial_rF(r_+,\theta)}}-\sqrt{(r-r_+)
\partial_rg(r_+,\theta)}(\partial_rW)\right]+mB=0,\nonumber\\\label{5}\\
&-&A\left[\sqrt{\frac{P\Omega^2(r_+,\theta)}{\rho^2(r_+,\theta)}}
(\partial_\theta W)\nonumber\right.\\
&+&\left.\frac{\iota\rho(r_+,\theta)
\Omega(r_+,\theta)}{\sqrt{\sin^2\theta
P[r^2+(a+l)^2]^2-Q(a\sin^2\theta+4l\sin^2\frac{\theta}{2})^2
}}\right.\nonumber\\&\times&\left.\left\{J-q \left[\frac{er[(l+a)^2-
(l^2+a^2\cos^2\theta+2la\cos\theta)]}{a[r^2+(l+a\cos\theta)^2]}
\right. \right.\right.\nonumber\\&+&\left.\left.\left.\frac{
g(l+a\cos\theta)[r^2+(l+a)^2]}{a[r^2+(l+a\cos\theta)^2]}\right]
\right\}\right]=0.\label{6}
\end{eqnarray}
The arbitrary function $W(r,\theta)$ can be separated as follows:
\cite{rbmann}
\begin{equation}
W(r,\theta)=R(r)+\Theta(\theta).\label{7}
\end{equation}

First we solve Eqs.(\ref{3})$-$(\ref{6}) for the massless case
($m=0$). Using the above separation, Eqs.(\ref{3}) and (\ref{5})
reduce to
\begin{eqnarray}
-B\left[\frac{-E+\Omega_HJ+\frac{qer}{[r_+^2+(a+l)^2]}}
{\sqrt{(r-r_+)\partial_rF(r_+,\theta)}}+
\sqrt{(r-r_+)\partial_rg(r_+,\theta)}R^\prime(r)\right]=0,\\
A\left[\frac{-E+\Omega_HJ+\frac{qer}{[r^2+(a+l)^2]}}
{\sqrt{(r-r_+)\partial_rF(r_+,\theta)}}-\sqrt{(r-r_+)
\partial_rg(r_+,\theta)}R^\prime(r)\right]=0,
\end{eqnarray}
implying that
\begin{eqnarray}
R^\prime(r)=R^\prime_+(r)&=&-R^\prime_-(r)=
\left[\frac{[r_+^2+(a+l)^2]}{(r-r_+)[1+\frac{\alpha(a-l)}{\omega}r_+]
[1-\frac{\alpha(a+l)}{\omega}r_+]}\right.\nonumber\\
&\times&\left.\frac{[E-\Omega_HJ-\frac{qer_+}
{[r_+^2+(a+l)^2]}]}{2[\frac{\alpha
l}{\omega}(\omega^2k+e^2+g^2)-M+\frac{\omega^2k}
{a^2-l^2}r_+]}\right],\label{AW}
\end{eqnarray}
where $R_+$ and $R_-$ correspond to the outgoing and incoming
solutions, respectively. This equation represents the pole at the
horizon, $r=r_+$.

In order to calculate the Hawking temperature by using the tunneling
approach, we need to regularize the singularity by specifying a
suitable complex contour to bypass the pole. In standard coordinate
representation, for outgoing particles (form inside of the horizon
to outside), we should take the contour to be an infinitesimal
semicircle below the pole $r=r_+$. Similarly, for the ingoing
particles (from outside to inside), the contour is above the pole.
Here, for the purpose of calculating the semiclassical tunneling
probability, we need to multiply the resulting wave equation by its
complex conjugate. In this way, the part of trajectory that starts
from outside of the BH and continues to the observer, will not
contribute to the calculation of the final tunneling probability and
can be ignored (since it will be entirely real). Therefore, the only
part of the wave equation (trajectory) that contributes to the
tunneling probability is the contour around the BH horizon. Instead
we choose a mathematically equivalent convention that the outgoing
contour is in the lower half plane and so do not multiply by a
negative sign \cite{rbmann}.

Integrating Eq.(\ref{AW}) around the pole, we obtain
\begin{eqnarray}
R_+(r)=-R_-(r)&=&\left[\frac{\pi
\iota[r_+^2+(a+l)^2]}{[1+\frac{\alpha(a-l)}{\omega}r_+]
[1-\frac{\alpha(a+l)}{\omega}r_+]}\right.\nonumber\\
&\times&\left.\frac{[E-\Omega_HJ-\frac{qer_+}
{[r_+^2+(a+l)^2]}]}{2[\frac{\alpha
l}{\omega}(\omega^2k+e^2+g^2)-M+\frac{\omega^2k}
{a^2-l^2}r_+]}\right].
\end{eqnarray}
The imaginary parts of $R_+$ and $R_-$ yield
\begin{eqnarray}
\textmd{Im}R_+=-\textmd{Im}R_-&=&\left[\frac{\pi[r_+^2+(a+l)^2]}
{[1+\frac{\alpha(a-l)}{\omega}r_+][1-\frac{\alpha(a+l)}{\omega}r_+]}
\right.\nonumber\\&\times&\left. \frac{[E-\Omega_HJ-\frac{qer_+}
{[r_+^2+(a+l)^2]}]}{2[\frac{\alpha
l}{\omega}(\omega^2k+e^2+g^2)-M+\frac{\omega^2k}
{a^2-l^2}r_+]}\right].
\end{eqnarray}
Thus the particle's tunneling probability from inside to outside the
horizon is
\begin{eqnarray}
\Gamma=\frac{\textmd{Prob}[\textmd{out}]}
{\textmd{Prob}[\textmd{in}]}&=&
\frac{\exp[-2(\textmd{Im}R_++\textmd{Im}\Theta)]}
{\exp[-2(\textmd{Im}R_-+\textmd{Im}\Theta)]}=
\exp[-4\textmd{Im}R_+]\nonumber\\
&=&\exp\left[\frac{-2\pi[r_+^2+(a+l)^2]}{[1+\frac{\alpha(a-l)}
{\omega}r_+][1-\frac{\alpha(a+l)}{\omega}r_+]}\nonumber\right.\\
&\times&\left.\frac{[E-\Omega_HJ-\frac{qer_+}{[r_+^2+(a+l)^2]}]}
{[\frac{\alpha l}{\omega}(\omega^2k+e^2+g^2)-M+\frac{\omega^2k}
{a^2-l^2}r_+]}\right].\label{12}
\end{eqnarray}

Here, $\Gamma$ is the tunneling probability for the classically
forbidden trajectories of the s-waves coming from inside to outside
the horizon. Using the WKB approximation, $\Gamma$ is given by
Eq.(1.1) in terms of classical action $I$ of Dirac particles
tunneling across the BH horizon as trajectories up to leading order
in $\hbar$. Thus, for calculating the Hawking temperature, we expand
the action in terms of particles energy $E$, i.e., $2I=\beta
E+O(E^2)$ so that the Hawking temperature is recovered at linear
order given by
\begin{equation}
\Gamma\sim\exp[-2I]\simeq\exp[-\beta E].
\end{equation}
This shows that the emission rate in the tunneling approach, up to
first order in $E$, retrieves the Boltzmann factor, $\exp[-\beta
E]$, where $\beta=\frac{1}{T_H}$ \cite{s2}. The higher-order terms
represent the self-interaction effects resulting from the energy
conservation.

Semiclassically, for the charged rotating BH solution, the energy
$E$ of the emitted particle in the Boltzmann factor should be
replaced by $E-\Omega_HJ$ due to the presence of the ergosphere. The
corresponding Hawking temperature at the event horizon takes the
form
\begin{equation}
T_H=\left[\frac{[\frac{\alpha
l}{\omega}(\omega^2k+e^2+g^2)-M+\frac{\omega^2k}
{a^2-l^2}r_+][1+\frac{\alpha(a-l)}{\omega}r_+]
[1-\frac{\alpha(a+l)}{\omega}r_+]}{2\pi[r_+^2+(a+l)^2]}\right]\label{NN5}.
\end{equation}
Thus, the Hawking temperature of Dirac particles tunneling from the
event horizon of the charged accelerating and rotating BHs with NUT
parameter is well described via fermions tunneling method. When
$l=0,~k=1$ in Eq.(\ref{NN5}), the Hawking temperature of the
accelerating and rotating BHs with electric and magnetic charges is
recovered \cite{ks1}. When $\alpha=0$, the Hawking temperature of
the charged accelerating and rotating BHs is reduced to the
temperature of non-accelerating BHs \cite{1-21}. For
$l=0,~k=1,~\alpha=0$ in Eq.(\ref{NN5}), the Hawking temperature of
the Kerr$-$Newman BH \cite{rbmann} is obtained which is further
reduced to the temperature of the RN BH (for $a=0$). Finally, in the
absence of charge, the temperature exactly reduces to the Hawking
temperature of the Schwarzschild BH \cite{A}.

For the massive case ($m\neq0$), following the same steps, we can
obtain the same temperature (\ref{NN5}). Thus the behavior of
massive particles near the BH horizon is the same as that for the
massless particles. For $l=0$, the tunneling probability (\ref{12})
reduces to the form for the pair of accelerating and rotating BHs
\cite{ks1}. For $\alpha=0$, we recover the tunneling probability for
the Kerr$-$Newman BH \cite{rbmann}. In the absence of rotation, this
yields the same result as for the RN BH \cite{rn}.

Now we explore the tunneling probability of charged massive and
massless fermions from the acceleration horizon $r_{\alpha_1}$ given
in Eq.(\ref{11}). The corresponding set of Eqs.(\ref{3})$-$(\ref{6})
for the outgoing and incoming fermions, respectively, yield
\begin{eqnarray}
R_+(r)&=&-R_-(r)=\left[\frac{\pi\iota[r_{\alpha_1}^2+(a+l)^2]}{\frac{\alpha}
{\omega}[l+\frac{\alpha}{\omega}(a^2-l^2)r_{\alpha_1}]}\right.\nonumber
\\&\times&\left.\frac{[E-\Omega_\alpha
J-\frac{qer_{\alpha_1}}{[r_{\alpha_1}^2+(a+l)^2]}] }{2[(\omega^2
k+e^2+g^2)(1+\frac{2\alpha
lr_{\alpha_1}}{\omega})-2Mr_{\alpha_1}+\frac{\omega^2k}{a^2-l^2}
r_{\alpha_1}^2]}\right].
\end{eqnarray}
The tunneling probability can be written as
\begin{eqnarray}
\Gamma&=&\exp\left[\frac{-2\pi[r_{\alpha_1}^2+(a+l)^2]}
{\frac{\alpha}{\omega}[l+\frac{\alpha}{\omega}(a^2-l^2)r_{\alpha_1}]}
\right.\nonumber\\&\times&\left.\frac{ [E-\Omega_\alpha
J-\frac{qer_{\alpha_1}}{[r_{\alpha_1}^2+(a+l)^2]}]}{
[(\omega^2k+e^2+g^2)(1+\frac{2\alpha lr_{\alpha_1}
}{\omega})-2Mr_{\alpha_1}+\frac{\omega^2kr_{\alpha_1}^2}
{a^2-l^2}]}\right].
\end{eqnarray}
Consequently, the Hawking temperature for the acceleration horizon
turns out to be
\begin{equation}
T_H=\left[\frac{\frac{\alpha}{\omega}[(\omega^2k+e^2+g^2)
(1+\frac{2\alpha
lr_{\alpha_1}}{\omega})-2Mr_{\alpha_1}+\frac{\omega^2kr_{\alpha_1}^2}
{a^2-l^2}][l+\frac{\alpha}{\omega}(a^2-l^2)r_{\alpha_1}]}
{2\pi[r_{\alpha_1}^2+(a+l)^2]}\right].
\end{equation}

\section{Action for the Emitted Particles}

In order to obtain the explicit expression for the action
$I_\uparrow$ in the spin-up case, we solve Eqs.(\ref{3})$-$(\ref{6})
near the BH horizon. Using Eq.(\ref{7}), we can write Eq.(\ref{3})
in the form
\begin{equation}
R^\prime(r)=\frac{mA}{B\sqrt{(r-r_+)
\partial_rg(r_+,\theta)}}-\frac{-E+\Omega_HJ+
\frac{qer_+}{[r_+^2+(a+l)^2]}}{(r-r_+)
\sqrt{\partial_rF(r_+,\theta)\partial_rg(r_+,\theta)}}.
\end{equation}
Integrating with respect to $r$, we get
\begin{eqnarray}
R(r)=R_+(r)=\int\frac{mA}{B\sqrt{(r-r_+)
\partial_rg(r_+,\theta)}}\textmd{d}r\nonumber\\
-\frac{\left(-E+\Omega_HJ+ \frac{qer_+}{[r_+^2+(a+l)^2]}\right)}{
\sqrt{\partial_rF(r_+,\theta)\partial_rg(r_+,\theta)}}\ln(r-r_+).
\end{eqnarray}
Similarly, for the incoming particles, Eq.(\ref{5}) yields
\begin{eqnarray}
R(r)=R_-(r)=\int\frac{mB}{A\sqrt{(r-r_+)
\partial_rg(r_+,\theta)}}\textmd{d}r\nonumber\\+
\frac{\left(-E+\Omega_HJ+\frac{qer_+}{[r_+^2+(a+l)^2]}\right)}{
\sqrt{\partial_rF(r_+,\theta)\partial_rg(r_+,\theta)}}\ln(r-r_+).
\end{eqnarray}
Using Eq.(\ref{7}), Eqs.(\ref{4}) and (\ref{6}) imply that
\begin{eqnarray}
&&\sqrt{\frac{P\Omega^2(r_+,\theta)}{\rho^2(r_+,\theta)}}
\partial_\theta\Theta+\frac{\iota\rho(r_+,\theta)
\Omega(r_+,\theta)}{\sqrt{\sin^2\theta
P[r_+^2+(a+l)^2]^2}}\nonumber\\&\times&\left[J
-q\left\{\frac{er_+[(l+a)^2-
(l^2+a^2\cos^2\theta+2la\cos\theta)]}{a[r_+^2+(l+a\cos\theta)^2]}
\nonumber\right.\right.\\&+&\left.\left.\frac{
g(l+a\cos\theta)[r_+^2+(l+a)^2]}{a[r_+^2+(l+a\cos\theta)^2]}
\right\}\right]=0.
\end{eqnarray}

Substituting the values of $\rho$ and $P$, after some manipulation,
it follows that
\begin{eqnarray}
\partial_\theta\Theta&=&\frac{\iota a^2\sin\theta J+ \iota
qer_+a\sin\theta}{[r_+^2+(a+l)^2]}\left[1-2\cos\theta\left\{\alpha
\frac{a}{\omega}M\right.\right.\nonumber\\&-&\left.\left.2\alpha^2l
\frac{a}{\omega^2}(\omega^2k+e^2+g^2)\right\}-\cos^2\theta\left\{-\alpha^2
\frac{a^2}{\omega^2}(\omega^2k+e^2+g^2)\right\}\right]^{-1}
\nonumber\\&+&\frac{-\iota aJ+\iota
qg(l+a\cos\theta)}{a\sin\theta}\left[1-2\cos\theta\left\{\alpha
\frac{a}{\omega}M\right.\right.\nonumber\\&-&\left.\left.2\alpha^2l
\frac{a}{\omega^2}(\omega^2k+e^2+g^2)\right\}-
\cos^2\theta\left\{-\alpha^2\frac{a^2}{\omega^2}(\omega^2k+e^2+g^2)\right\}\right]^{-1}
\nonumber\\&+&\frac{2la\iota (1-\cos\theta)J+\iota qer_+
2l(1-\cos\theta)}{\sin\theta[r_+^2+(a+l)^2]}\left[1-2\cos\theta
\left\{\alpha\frac{a}{\omega}M\right.\right.\nonumber\\&-&\left.\left.2\alpha^2l
\frac{a}{\omega^2}(\omega^2k+e^2+g^2)\right\}-\cos^2\theta\left\{-\alpha^2
\frac{a^2}{\omega^2}(\omega^2k+e^2+g^2)\right\}\right]^{-1}.\nonumber\\
\end{eqnarray}
Integrating with respect to $\theta$, we have
\begin{equation}
\Theta=\frac{\iota[qaer_++Ja^2]}{[r_+^2+(a+l)^2]}I_1+I_2+
\frac{2l\iota[aJ+qer_+]}{[r_+^2+(a+l)^2]}I_3,\label{10}
\end{equation}
where $I_1,~I_2$ and $I_3$ are given as follows
\begin{eqnarray}
I_1&=&\int\sin\theta\left[
1-2\cos\theta\left\{\alpha\frac{a}{\omega}M-2\alpha^2l
\frac{a}{\omega^2}(\omega^2k+e^2+g^2)\right\}\right.\nonumber\\&-&\left.\cos^2\theta\left\{-\alpha^2
\frac{a^2}{\omega^2}(\omega^2k+e^2+g^2)\right\}\right]^{-1}\textmd{d}\theta,\\
I_2&=&\int\left[\frac{\iota \{q g(l+a\cos\theta)-a
J\}}{a\sin\theta}\right]\left[1-2\cos\theta
\left\{\alpha\frac{a}{\omega}M\right.\right.\nonumber\\&-&\left.\left.2\alpha^2l
\frac{a}{\omega^2}(\omega^2k+e^2+g^2)\right\}-\cos^2\theta\left\{-\alpha^2
\frac{a^2}{\omega^2}(\omega^2k+e^2+g^2)\right\}\right]^{-1}\textmd{d}\theta,\nonumber\\\\
I_3&=&\int\left[\frac{1-\cos\theta}{\sin\theta}\right]
\left[1-2\cos\theta\left\{\alpha\frac{a}{\omega}M-2\alpha^2l
\frac{a}{\omega^2}(\omega^2k+e^2+g^2)\right\}\right.\nonumber\\&-&\left.\cos^2\theta\left\{-\alpha^2
\frac{a^2}{\omega^2}(\omega^2k+e^2+g^2)\right\}\right]^{-1}\textmd{d}\theta.
\end{eqnarray}
Solving these integrals, we obtain after some algebra
\begin{eqnarray}
I_1&=&\left[\frac{1}{2\alpha\frac{a}{\omega}\sqrt{\left\{M-2\alpha\frac{l}{\omega}(\omega^2k+e^2+g^2)\right\}^2-
(\sqrt{\omega^2k+e^2+g^2})^2}}\right]\nonumber\\&\times&
\ln\left[\left(1-\alpha\frac{a}{\omega}\cos\theta\left\{M-2\alpha
\frac{l}{\omega}(\omega^2k+e^2+g^2)\right.\right.\right.\nonumber\\&+&\left.\left.
\sqrt{\left\{M-2\alpha\frac{l}{\omega}(\omega^2k+e^2+g^2)\right\}^2-
(\sqrt{\omega^2k+e^2+g^2})^2}\right\}\right)\nonumber\\&\times&
\left(1-\alpha\frac{a}{\omega}\cos\theta\left\{M-2\alpha\frac{l}{\omega}(\omega^2k+e^2+g^2)
\right.\right.\nonumber\\&-&\left.\left.\left.
\sqrt{\left\{M-2\alpha\frac{l}{\omega}(\omega^2k+e^2+g^2)\right\}^2-
(\sqrt{\omega^2k+e^2+g^2})^2}\right\}\right)^{-1}\right],\nonumber\\
\end{eqnarray}
\begin{eqnarray}
I_2&=&L_1\ln\left[\left(1-\alpha\frac{a}{\omega}\cos\theta\left\{M-2\alpha\frac{l}{\omega}
(\omega^2k+e^2+g^2)\right.\right.\right.\nonumber\\&+&\left.\left.
\sqrt{\left\{M-2\alpha\frac{l}{\omega}(\omega^2k+e^2+g^2)\right\}^2-
(\sqrt{\omega^2k+e^2+g^2})^2}\right\}\right)\nonumber\\&\times&
\left(1-\alpha\frac{a}{\omega}\cos\theta\left\{M-2\alpha\frac{l}{\omega}
(\omega^2k+e^2+g^2)\right.\right.\nonumber\\&-&\left.\left.\left.
\sqrt{\left\{M-2\alpha\frac{l}{\omega}(\omega^2k+e^2+g^2)\right\}^2-
(\sqrt{\omega^2k+e^2+g^2})^2}\right\}\right)^{-1}\right]\nonumber\\&+&L_2
\ln\left[1-2\cos\theta\left\{\alpha\frac{a}{\omega}M-2\alpha^2l\frac{a}{\omega^2}
(\omega^2k+e^2+g^2)\right\}\right.\nonumber\\&+&\left.\alpha^2
\frac{a^2}{\omega^2}{\cos^2\theta}(\omega^2k+e^2+g^2)\right]+L_3
\ln[1-\cos\theta]\nonumber\\&+&L_4\ln[1+\cos\theta],
\end{eqnarray}
where
\begin{eqnarray}
L_1&=&\left[\left\{2\alpha\frac{a}{\omega}\sqrt{\left\{M-2\alpha\frac{l}{\omega}
(\omega^2k+e^2+g^2)\right\}^2-(\sqrt{\omega^2k+e^2+g^2})^2}\right\}\right.\nonumber\\
&\times&\left\{\left(1+\alpha^2\frac{a^2}{\omega^2}(\omega^2k+e^2+g^2)\right)^2-
4\alpha^2\frac{a^2}{\omega^2}\right.\nonumber\\&\times&\left.\left.
\left(M-2\alpha\frac{l}{\omega}(\omega^2k+e^2+g^2)\right)^2\right\}\right]^{-1}
\left[\iota
J\left\{2\left(\alpha\frac{a}{\omega}M\right.\right.\right.\nonumber\\&-&\left.
\left.\left.2\alpha^2l\frac{a}{\omega^2}(\omega^2k+e^2+g^2)\right)^2-
\alpha^4\frac{a^4}{\omega^4}(\omega^2k+e^2+g^2)^2\right.\right.\nonumber\\&-&
\left.\alpha^2\frac{a^2}{\omega^2}(\omega^2k+e^2+g^2)\right\}+ \iota
q
g\left\{-\alpha\frac{a}{\omega}M+2\alpha^2l\frac{a}{\omega^2}(\omega^2k+e^2+g^2)
\right.\nonumber\\&+&
\frac{\alpha^4l}{a}\frac{a^4}{\omega^4}(\omega^2k+e^2+g^2)^2-
\frac{2\alpha^2l}{a}\frac{a^2}{\omega^2}\left(M-2\alpha\frac{l}{\omega}
(\omega^2k+e^2+g^2)\right)^2\nonumber\\&+& \frac{\alpha^2l}{a}
\frac{a^2}{\omega^2}(\omega^2k+e^2+g^2)+\alpha^3\frac{a^3}{\omega^3}(\omega^2k+e^2+g^2)
\nonumber\\&\times&\left.\left.\left(M-2\alpha\frac{l}{\omega}(\omega^2k+e^2+g^2)\right)
\right\}\right],
\end{eqnarray}
\begin{eqnarray}
L_2&=&\left[\left\{1+\alpha^2\frac{a^2}{\omega^2}(\omega^2k+e^2+g^2)\right\}^2-
4\alpha^2\frac{a^2}{\omega^2}\right.\nonumber\\&\times&
\left.\left\{M-2\alpha\frac{l}{\omega}\left(\omega^2k
+e^2+g^2\right)\right\}^2\right]^{-1}\nonumber\\&\times&\left[\iota
J\left\{\alpha\frac{a}{\omega}M-2\alpha^2l\frac{a}{\omega^2}(\omega^2k+e^2+g^2)\right\}\right.
\nonumber\\&-&\frac{\iota
qg}{2}\left\{2\frac{l}{a}\left(\alpha\frac{a}{\omega}M-2\alpha^2l
\frac{a}{\omega^2}(\omega^2k+e^2+g^2)\right)\right.\nonumber\\&+&\left.\left.1+
\alpha^2\frac{a^2}{\omega^2}(\omega^2k+e^2+g^2)\right\}\right],\\
L_3&=&\frac{1}{2}\left[\iota qg\left(\frac{l}{a}+1\right)-\iota
J\right]\left[1-2\left\{\alpha\frac{a}{\omega}M-2\alpha^2l\right.\right.\nonumber\\
&\times&\left.\left.\frac{a}{\omega^2}
(\omega^2k+e^2+g^2)\right\}+\alpha^2\frac{a^2}{\omega^2}(\omega^2k+e^2+g^2)\right]^{-1},\\
L_4&=&\frac{1}{2}\left[\iota J+\iota qg
\left(1-\frac{l}{a}\right)\right]\left[1+\alpha^2\frac{a^2}{\omega^2}(\omega^2k+e^2+g^2)\right.
\nonumber\\&+&\left.2\left\{\alpha\frac{a}{\omega}M-2\alpha^2l\frac{a}{\omega^2}
(\omega^2k+e^2+g^2)\right\}\right]^{-1}
\end{eqnarray}
and $I_3$ can be obtained as
\begin{eqnarray}
I_3&=&N_1\ln\left[\left(1-\alpha\frac{a}{\omega}\cos\theta\left\{M-2\alpha\frac{l}{\omega}
(\omega^2k+e^2+g^2)\right.\right.\right.\nonumber\\&+&\left.\left.
\sqrt{\left\{M-2\alpha\frac{l}{\omega}(\omega^2k+e^2+g^2)\right\}^2-(\sqrt{\omega^2k+e^2+g^2})^2}
\right\}\right)\nonumber\\&\times&\left(1-\alpha\frac{a}{\omega}\cos\theta\left\{M-2\alpha\frac{l}{\omega}
(\omega^2k+e^2+g^2)\right.\right.\nonumber\\&-&\left.\left.\left.\sqrt{\left\{M-2\alpha\frac{l}{\omega}
(\omega^2k+e^2+g^2)\right\}^2-(\sqrt{\omega^2k+e^2+g^2})^2}\right\}\right)^{-1}\right]
\nonumber\\&+&N_2\ln\left[1-2\cos\theta\left\{\alpha\frac{a}{\omega}M-2\alpha^2l\frac{a}{\omega^2}
(\omega^2k+e^2+g^2)\right\}\right.\nonumber\\&+&\left.\alpha^2\frac{a^2}{\omega^2}{\cos^2\theta}
(\omega^2k+e^2+g^2)\right]+N_3\ln[1+\cos\theta],
\end{eqnarray}
where
\begin{eqnarray}
N_1&=&\left[\frac{\alpha\frac{
a}{\omega}M-2l\alpha^2\frac{a}{\omega^2}(\omega^2k+e^2+g^2)+\alpha^2\frac{a^2}{\omega^2}
(\omega^2k+e^2+g^2)}{2\alpha\frac{a}{\omega}\sqrt{\left\{M-2\alpha\frac{l}{\omega}
(\omega^2k+e^2+g^2)\right\}^2-(\sqrt{\omega^2k+e^2+g^2})^2}}\right]
\nonumber\\&\times&\left[1+\alpha^2\frac{a^2}{\omega^2}(\omega^2k+e^2+g^2)+2\alpha
\left\{\frac{a}{\omega}M\right.\right.\nonumber\\&-&\left.\left.2\alpha^2l\frac{a}{\omega^2}
(\omega^2k+e^2+g^2)\right\}\right]^{-1},\\
N_2&=&\frac{1}{2}\left[1+\alpha^2\frac{a^2}{\omega^2}(\omega^2k+e^2+g^2)+2
\left\{\alpha\frac{a}{\omega}M\right.\right.\nonumber\\&-&\left.\left.2\alpha^2l\frac{a}{\omega^2}
(\omega^2+e^2+g^2)\right\}\right]^{-1},\\
N_3&=&-\left[1+\alpha^2\frac{a^2}{\omega^2}(\omega^2k+e^2+g^2)+2
\left\{\alpha\frac{a}{\omega}M\right.\right.\nonumber\\&-&\left.\left.2\alpha^2l\frac{a}{\omega^2}
(\omega^2k+e^2+g^2)\right\}\right]^{-1}.
\end{eqnarray}
Inserting these values of the integrals in Eq.(\ref{10}) and using
Eq.(\ref{7}), we obtain $W(r,\theta)$. This leads one to evaluate
the action for the outgoing massive particles. For $m=0$, this
reduces to the action for the massless particles. Similarly, we can
determine the action for the incoming massive and massless
particles.

\section{Outlook}

In this paper, we have used semiclassical WKB approximation to study
tunneling of charged fermions from a pair of accelerating and
rotating BHs having electric and magnetic charges, together with a
NUT parameter. Classically, a particle can only fall inside the
horizon while in the semiclassical approach, the horizon plays a
role of two way energy barrier for a pair of positive and negative
energy particles. The positive energy particles have ability to
tunnel outside the event horizon, which contradicts classical
approach. Thus, we have considered tunneling probabilities for both
incoming as well as outgoing particles. Relating these tunneling
probabilities with the Boltzmann factor $\exp[-\beta E]$ for
emission at the Hawking temperature, we can recover the
corresponding Hawking temperature for this pair of BHs at event
horizons.

The tunneling probabilities of outgoing/incoming charged fermions do
not depend upon mass of the fermions but only its charge. The
corresponding Hawking temperature depends upon mass, acceleration
and rotation parameters and also NUT parameter as well as electric
and magnetic charges of the pair of BHs. Equations for the spin-down
case are of the same form as for the spin-up case except for a
negative sign. For both massive and massless cases, the Hawking
temperature implies that both spin-up and spin-down particles are
emitted at the same rate \cite{rbmann1}.

It is worth mentioning here that in the absence of the NUT parameter
$l=0$, all the results reduce to the results of the pair of charged
accelerating and rotating BHs \cite{ks1}. Further, $\alpha=0$
provides the results of the Kerr$-$Newman BH \cite{rbmann} and $a=0$
yields the results of the RN BH \cite{rn}.

\vspace{0.25cm}

{\bf Acknowledgement}

\vspace{0.25cm}

We would like to thank the Higher Education Commission, Islamabad,
Pakistan, for its financial support through the {\it Indigenous
Ph.D. 5000 Fellowship Program Batch-IV}.


\begin{thebibliography}{99}

\bibitem{hawk1} Hawking, S.W.: Nature {\bf 248}(1974)30; Commun.
Math. Phys. {\bf 43}(1975)199.

\bibitem{parwil} Parikh, M.K. and Wilczek, F.: Phys. Rev.
Lett. {\bf 85}(2000)5042.

\bibitem{krawil} Kraus, P. and Wilczek, F.: Nucl.
Phys. {\bf B433}(1995)403.

\bibitem{abhen} Angheben, M., Nadalini, M., Vanzo, L. and Zerbini,
S.: JHEP {\bf 05}(2005)014.

\bibitem{Srini1} Srinivasan, K. and Padmanabhan,
T.: Phys. Rev. {\bf D60}(1999)024007.

\bibitem{Srini2} Shankaranarayanan, S., Srinivasan, K.
and Padmanabhan, T.: Mod. Phys. Lett.
{\bf A16}(2001)571; \emph{ibid}. Class. Quantum Grav.
{\bf19}(2002)2671.

\bibitem{WKB} Brillouin, L.: Comptes Rendus de l'Academie des Sciences {\bf
183}(1926)24; Kramers, H.A.: Zeitschrift der Physik {\bf
39}(1926)828; Wentzel, G.: Zeitschrift der Physik  {\bf
38}(1926)518.

\bibitem{23-1} Page, D.N.: Phys. Rev. {\bf D13}(1976)198.

\bibitem{23-2} Page, D.N.: Phys. Rev. {\bf D14}(1976)3260.

\bibitem{rbmann1} Kerner, R. and Mann, R.B.: Class. Quantum Grav. {\bf
25}(2008)095014.

\bibitem{rbmann} Kerner, R. and Mann, R.B.: Phys. Lett. {\bf B665}(2008)277.

\bibitem{ks1} Gillani, U.A. and Saifullah, K.: Phys. Lett. {\bf B699}(2011)15;
Gillani, U.A., Rehman, M. and Saifullah, K.: JCAP {\bf 06}(2011)016;
Rehman, M. and Saifullah, K.: JCAP {\bf 03}(2011)001.

\bibitem{1-21} Bilal, M. and Saifullah, K.: arXiv:1010.5575.

\bibitem{S1} Sharif, M. and Javed, W.: J. Korean Phys. Soc. {\bf
57}(2010)217; Astrophys. Space Sci. \textbf{337}(2012)335; JETP
\textbf{141}(2012)1071; \emph{Charged Fermions Tunneling from
Regular Black Holes} JETP (2012, to appear).

\bibitem{s2} Kerner, R. and Mann, R.B.: Phys. Rev. {\bf D73}(2006)104010.

\bibitem{[1]} Griffiths, J.B.: \emph{Colliding Plane Waves in General Relativity}
(Oxford University Press, 1991).

\bibitem{[8]} Plebanski, J.F. and Demianski, M.: Ann. Phys. {\bf 98}(1976)98.

\bibitem{[9]} Chen, W., L$\ddot{u}$, H. and Pope, C.N.: Class. Quantum Grav. {\bf
23}(2006)5323; Nucl. Phys. {\bf B762}(2007)38.

\bibitem{[11]} Griffiths, J.B. and Padolsky, J.: Class. Quantum Grav. {\bf
24}(2007)1687.

\bibitem{[12]} Al-Badawi, A. and Halilsoy, M.: Gen. Relativ. Gravit. {\bf
38}(2006)1729.

\bibitem{acc} Griffiths, J.B. and  Podolsk$\acute{y}$, J.:
Class. Quantum Grav. {\bf 22}(2005)3467.

\bibitem{NUT} Newmann, E.T., Tamburino, L.A. and Unti, T.: J. Math.
Phys. {\bf 4}(1963)915.

\bibitem{acc1} Podolsk$\acute{y}$, J. and Kadlecov$\acute{a}$,
H.: Class. Quantum Grav. {\bf 26}(2009)105007.

\bibitem{rn} Zeng, X.X. and Yang, S.Z.: Gen. Relativ. Gravit. {\bf
40}(2008)2107.

\bibitem{A} Chen, D.Y., Jiang, Q.Q.. and Zua, X.T.: Phys. Lett. {\bf B665}(2008)106.

\end{thebibliography}
\end{document}